\documentclass[conference]{IEEEtran}
\IEEEoverridecommandlockouts

\usepackage{silence}
\usepackage{cite}
\usepackage{amsmath,amssymb,amsfonts}
\usepackage{algorithmic}
\usepackage{graphicx}
\usepackage{textcomp}
\usepackage{xcolor}
\usepackage{subcaption}
\usepackage{multirow}
\usepackage{booktabs}

\newcommand{\bb}[1]{\mathbf{#1}}

\def\BibTeX{{\rm B\kern-.05em{\sc i\kern-.025em b}\kern-.08em
    T\kern-.1667em\lower.7ex\hbox{E}\kern-.125emX}}
\begin{document}



\title{Calibration of Multiple Asynchronous Microphone Arrays using Hybrid TDOA}

\author{Chengjie Zhang, Wenda Pan, Xinyang Han, and He Kong\thanks{Chengjie Zhang and Wenda Pan contributed equally to this work. This work was supported by the Science, Technology, and Innovation Commission of 
Shenzhen Municipality, China, under Grant No. ZDSYS20220330161800001, the Shenzhen Science and Technology Program under Grant No. KQTD20221101093557010, the Guangdong Science and Technology Program (Grant No. 2024B1212010002), and the National Natural Science Foundation of China 
under Grant No. 62350055. The authors are with the Shenzhen Key Laboratory of Control Theory and Intelligent Systems, and the Guangdong Provincial Key Laboratory of Fully Actuated System Control Theory and Technology, at the Southern University of Science and Technology, Shenzhen 518055, China. Emails: [12332644,12111907,12112608]@mail.sustech.edu.cn; kongh@sustech.edu.cn.}}

\maketitle

\begin{abstract}
Accurate calibration of acoustic sensing systems made of multiple asynchronous microphone arrays is essential for satisfactory performance in sound source localization and tracking. State-of-the-art calibration methods for this type of system rely on the time difference of arrival and direction of arrival measurements among the microphone arrays (denoted as TDOA-M and DOA, respectively). In this paper, to enhance calibration accuracy, we propose to incorporate the time difference of arrival measurements between adjacent sound events (TDOA-S) with respect to the microphone arrays. More specifically, we propose a two-stage calibration approach, including an initial value estimation (IVE) procedure and the final joint optimization step. The IVE stage first initializes all parameters except for microphone array orientations, using hybrid TDOA (i.e., TDOA-M and TDOA-S), odometer data from a moving robot carrying a speaker, and DOA. Subsequently, microphone orientations are estimated through the iterative closest point method. The final joint optimization step estimates multiple microphone array locations, orientations, time offsets, clock drift rates, and sound source locations simultaneously. Both simulation and experiment results show that for scenarios with low or moderate TDOA noise levels, our approach outperforms existing methods in terms of accuracy. All code and data are available at https://github.com/AISLAB-sustech/Hybrid-TDOA-Multi-Calib.        
\end{abstract}


\section{Introduction}
Microphone arrays are widely used in practice for sound source localization/tracking \cite{app-track, app-uav,evers2020locata,lagace2023ego}, acoustic mapping \cite{Evers2018,hu2023, Fu2024}, environmental monitoring \cite{Grondin,soundLocReview,app-nakadai}, multi-modal fusion \cite{Molina,Manocha2017,qian2021multi,Verellen2020,Ferreira}, etc. For satisfactory performance, it is essential to accurately calibrate these systems \cite{WangLin, SuKong,su2015simultaneous,su2020asynchronous, Li2024}. In particular, calibration of multiple microphone arrays is more challenging than doing so for a single array \cite{Ono, Plinge2016}, as one needs to estimate relative positions and orientations, and potentially the asynchronous factors between arrays. Earlier works on the calibration of multiple microphone arrays mostly focus on the 2D scenario using inter-array time difference of arrival (TDOA) and direction of arrival (DOA) \cite{2D-calib1, 2D-calib2, 2D-Calib3}. 

Only a few works have considered the multiple array geometric calibration problem in 3D \cite{3D-Calib1,3D-Calib2, 3D-Calib3, wang}. For example, by assuming that at least three sound source positions are known, \cite{3D-Calib1} has considered calibration of multiple arrays in 3D solely based on DOA. The methods in \cite{3D-Calib2, 3D-Calib3} can simultaneously estimate asynchronous array position, orientation and time offset parameters, and sound source positions, but they do not consider the clock drift rate (caused by the sampling rate mismatch between arrays). 

To overcome the above limitations, in our recent work \cite{wang}, by using TDOA, DOA, and odometer information (relative position measurement of the robot that carries a sound source and moves in the environment), for the 3D case, we have proposed a batch optimization-based calibration framework to simultaneously estimate the position, rotation, time offset, and clock drift rate between microphone arrays and sound source locations.

\begin{figure}[tp]
\centering
\includegraphics[scale=0.1]{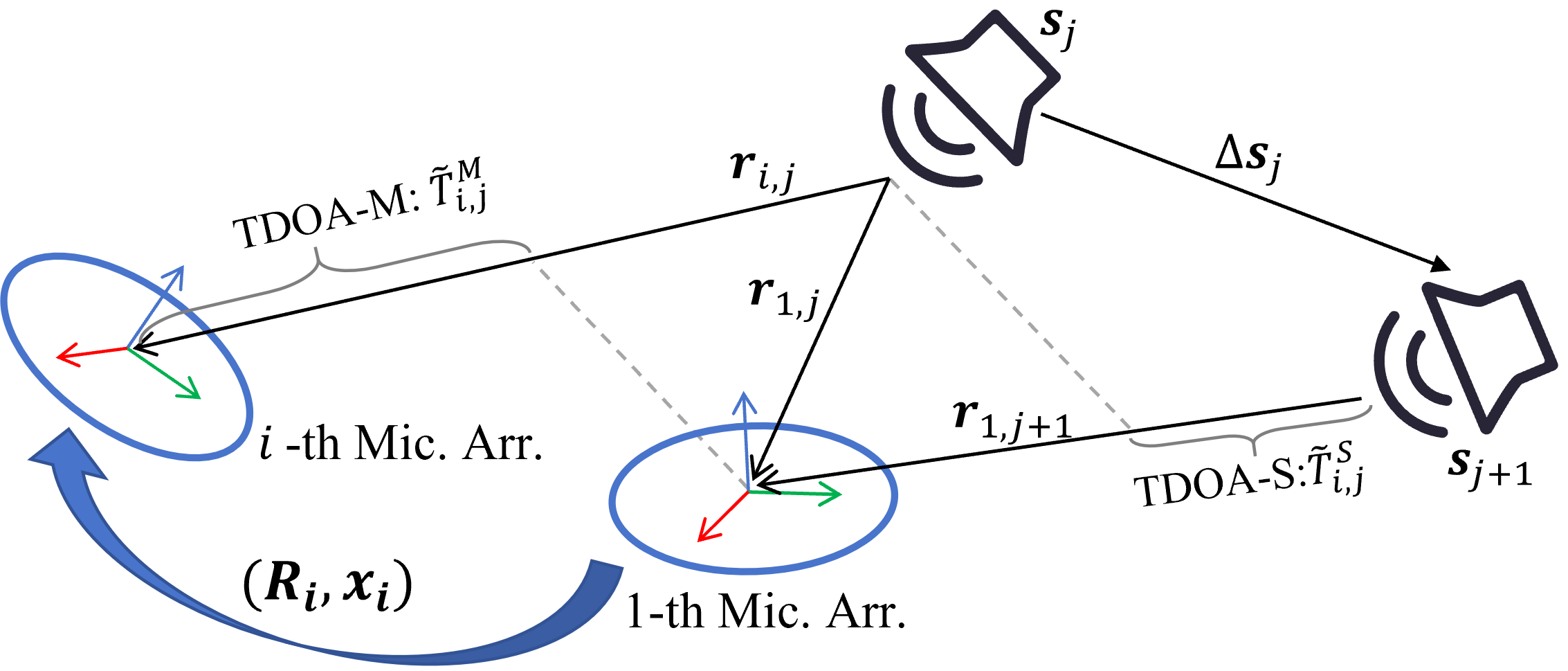}
\caption{The calibration scenario for multiple microphone arrays using hybrid TDOA information. Two microphone arrays (the 1-th and i-th) and two sound events (the j-th and j+1-th) are used for illustration here.}
\label{fig:calibScene}
\end{figure}

In parallel, very recently, for calibrating single microphone arrays \cite{Hybrid-TDOA}, we have introduced a framework that incorporates the time difference of arrival measurements between adjacent sound events (TDOA-S) with respect to (w.r.t.) a certain microphone, in addition to the traditional time difference of arrival measurements among the microphones (TDOA-M). It has been shown in \cite{Hybrid-TDOA} that using hybrid TDOA information (i.e., both TDOA-S and TDOA-M) leads to improved calibration results compared to the case using only TDOA-M information.

Motivated by the ideas in \cite{Hybrid-TDOA}, in this paper, we propose a calibration framework for multiple asynchronous microphone arrays using hybrid TDOA (see in Fig. \ref{fig:calibScene}), to enhance the calibration accuracy. Our proposed approach is comprised of an initial value estimation (IVE) procedure and a final joint optimization step. The IVE stage firstly initializes all parameters except for microphone array orientations using hybrid TDOA (i.e., TDOA-M and TDOA-S), odometer data from a moving robot carrying a speaker, and DOA of the first microphone. Secondly, each microphone array's rotation is estimated through iterative closest point (ICP) method. The final joint optimization step estimates multiple microphone array locations, orientations, time offsets, clock drift rates, and sound source locations simultaneously. Extensive simulations and experiments show that for scenarios with low or moderate TDOA noise levels, our approach results in improved accuracy, in comparison to \cite{wang}.

\section{Preliminaries and Problem Formulation}  
\label{sec:format}
As shown in Fig. \ref{fig:calibScene}, in our calibration scenario, there are $N$ microphone arrays and $K$ sound events. Denote the $i$-th microphone array location, orientation, time offset, and clock drift rate as $\bb{x}_i$, $\bb{\phi}_i$, $\tau_i$, and $\delta_i$, respectively, for \(i = 1, 2, \ldots, N\). $\bb{\phi}_i$ stands for the rotation of $i$-th microphone array in the global coordinate frame, i.e., $Mic.$ frame. We use Lie algebra \cite[Chp. 7]{stateEstRobotics}, \cite[Chp. 3]{lieGroup} to represent $\bb{\phi}_i$, i.e., $\bb{\phi}_i\in\mathfrak{so}(3), ||\bb{\phi}_i||_2\leq\pi$. The $j$-th sound event location and emitting time are denoted as $\bb{s}_j$ and $t_j$, respectively, where \(j = 1, 2, \ldots, K\).


Without loss of generality, the first microphone array frame is taken to be $Mic.$ frame, with $\bb{x}_1=\bb{0}$ and $\bb{\phi}_1=\bb{0}$. Additionally, we assume that $t_1=0$. As in existing works (see \cite{wang} and the references therein), we assume that the geometry of individual arrays is known and each channel in an array is synchronous. Parameters to be estimated from the measurements are $\bb{x}_{mic}=[\delta_1,\bb{x}_2,\bb{\phi}_2,\tau_{2,1},\delta_2, ...,\bb{x}_N,\bb{\phi}_N,\tau_{N,1},\delta_N]^T$ and $\bb{s}=[\bb{s}_1,\bb{s}_2,...,\bb{s}_K]^T$, where $\tau_{i,1}=\tau_i-\tau_1$, $i>1$. 

Following \cite{Hybrid-TDOA}, we use inter-array hybrid TDOA for estimating the aforementioned unknown parameters. More specifically, assume $\bb{d}_{i,j}=\bb{x}_i-\bb{s}_j$ and $d_{i,j}=||\bb{d}_{i,j}||_2$. Then the measurement model of TDOA-S is
\begin{equation}
\label{eq:tdoa-s}
T^{S}_{i,j}=\underbrace{\frac{d_{i,j+1}-d_{i,j}}{c}}_{\widetilde{T}^{S}_{i,j}}+(1+\delta_i)\Delta t_j,
\end{equation}
where $j<K$ and $\Delta t_j=t_{j+1}-t_j$ denotes the time interval between consecutive sound events. The measurement model of TDOA-M is 
\begin{equation}
\label{eq:tdoa-m}
    T^{M}_{i,j}=\underbrace{\frac{d_{i,j}-d_{1,j}}{c}}_{\widetilde{T}^{M}_{i,j}}+\tau_{i,1}+\delta_{i,1}t_{j,1},
\end{equation}
where $\delta_{i,1}=\delta_i-\delta_1$ ($i>1$). For $j>1$, $t_{j,1}=t_j-t_1=\sum^{j}_{k=2}\Delta t_{k-1}$. Detailed procedures for extracting TDOA-S can be found in \cite{Hybrid-TDOA}. The measurement model of DOA is 
\begin{equation}
\label{eq:doa}
    \bb{r}_{i,j}=\bb{R}_i^T\frac{\bb{d}_{i,j}}{d_{i,j}},
\end{equation}
where $\bb{R}_i$ is the rotation matrix that rotates a vector in the coordinate frame of $i$-th microphone array to $Mic.$ frame. $\bb{R}_i$ can be converted from Lie algebra with the following formula 
\cite[Chp. 7]{stateEstRobotics}, \cite[Chp. 3]{lieGroup}: $\bb{R}_i=cos(\theta_i)\bb{I}+(1-cos(\theta_i))\bb{n}_i\bb{n}_i^T+sin(\theta_i)\bb{n}_i^\wedge$ where $\bb{\phi}_i=\theta_i\bb{n}_i$ and $||\bb{n}_i||_2=1$.

Hence, without noise, the total hybrid TDOA measurements $\bb{T}^H\in\mathbb{R}^{(N-1)K+N(K-1)\times1}$ are
\begin{equation}
    \bb{T}^H=[\bb{T}^S,\bb{T}^M]^T,
\end{equation}
where $\bb{T}^S=[\bb{T}^S_1,\bb{T}^S_2,...,\bb{T}^S_N]^T$, $\bb{T}^S_i=[T^S_{i,1},T^S_{i,2},...,T^S_{i,K-1}\\]^T$ and $\bb{T}^M=[\bb{T}^M_1,\bb{T}^M_2,...,\bb{T}^M_K]^T$, $\bb{T}^M_j=[T^M_{2,j},T^M_{3,j},...,\\T^M_{N,j}]^T$,
and the total DOA measurements $\bb{R}^D \in \mathbb{R}^{3NK\times1}$ are
\begin{equation}
    \bb{r}=[\bb{r}_{1,1},\bb{r}_{2,1},...,\bb{r}_{N,1},\bb{r}_{1,2},\bb{r}_{2,2},...,\bb{r}_{N,2},...,\bb{r}_{N,K}]^T.
\end{equation}
Considering i.i.d Gaussian noises, the real TDOA-M and TDOA-S measurements are $t^{M}_{i,j}=T^{M}_{i,j}+w^{M}_{i,j}$ ($i>1$) and  $t^{S}_{i,j}=T^{S}_{i,j}+w^{S}_{i,j}$ ($j<K$), respectively, with $w^{M}_{i,j},w^{S}_{i,j}\sim N(0,\sigma^2_{tdoa})$. The real hybrid TDOA measurements are
\begin{equation}
    \bb{t}^H=[\bb{t}^S,\bb{t}^M]^T,
\end{equation}
where $\bb{t}^S=[\bb{t}^S_1,\bb{t}^S_2,...,\bb{t}^S_N]^T$, $\bb{t}^S_i=[t^S_{i,1},t^S_{i,2},...,t^S_{i,K-1}]^T$ and $\bb{t}^M=[\bb{t}^M_1,\bb{t}^M_2,...,\bb{t}^M_K]^T$, $\bb{t}^M_j=[t^M_{2,j},t^M_{3,j},...,t^M_{N,j}]^T$. 


When using Lie algebra to represent the relative orientation between the microphone arrays, the real DOA measurement is $\bb{r}_{i,j}^D= \bb{R}^e_{i,j}\bb{R}_i^T\frac{\bb{d}_{i,j}} {d_{i,j}}$ 
where $\bb{R}^e_{i,j}$ stands for a multiplicative noise term \cite[Chp. 7]{stateEstRobotics}. Then the entire real DOA measurements are
\begin{equation}
    \bb{r}^D=[\bb{r}^D_{1,1},\bb{r}^D_{2,1},...,\bb{r}^D_{N,1},\bb{r}^D_{1,2},\bb{r}^D_{2,2},...,\bb{r}^D_{N,2},...,\bb{r}^D_{N,K}]^T.
\end{equation}
Under Gaussian noise $\bb{v}_j\sim N(\bb{0},\sigma^2_{odo}\bb{I}_3)$, the odometer measurements are $\bb{m}=[\bb{m}_1,\bb{m}_2,..., \bb{m}_{K-1}]^T$ with $\bb{m}_j$ being defined as follows
\begin{equation}
\label{eq:odo}
    \bb{m}_j=\Delta\bb{s}_j+\bb{v}_j=\bb{s}_{j+1}-\bb{s}_j+\bb{v}_j,
\end{equation}
where $j<K$. For later use, denote $\Delta\bb{s}=[\Delta\bb{s}_1,\Delta\bb{s}_2, ..., \Delta\bb{s}_{K-1}]^T$.

\section{The Proposed Approach}
Our approach includes an IVE procedure and the final joint optimization step, as to be described below.
\subsubsection{Initial Value Estimation} 
The IVE procedure has two steps. Firstly, we construct a weighted NLS based on the DOA of the first microphone array, hybrid TDOA, and odometer measurements and solve it using the Gauss-Newton (GN) method to estimate the locations of the microphone array, time offsets, clock drift rates, and sound source locations simultaneously. Secondly, based on the estimated microphone positions, estimated sound source positions, and DOA, we use the ICP method \cite{ICP,icp_sovler} to estimate microphone array orientations.

For the first step, we generalize the method \cite{Hybrid-TDOA} from the single microphone array case to the multiple array scenario (a microphone array in the latter case is conceptually a microphone in the former case). Since the location and orientation of the first microphone array are fixed, the corresponding DOA only involves sound source position parameters, which can be incorporated into the NLS to improve calibration accuracy further. Because the formulation of the NLS is similar to the NLS in final joint optimization Eq. (\ref{NLS}), mathematical expressions of the NLS are omitted here.

For the second step, given microphone array locations $\hat{\bb{x}}_2, \hat{\bb{x}}_3,...,\hat{\bb{x}}_N$ and sound source locations $\hat{\bb{s}}_1,\hat{\bb{s}}_2,...,\hat{\bb{s}}_K$ estimated from the first step in our IVE method and DOA measurements, we compute $j$-th sound source location in the coordinate frame established by $i$-th microphone array ($i>1$) via
\begin{equation}
    \hat{\bb{s}}^i_j=||\hat{\bb{x}}_i-\hat{\bb{s}}_j||\bb{r}_{i,j}.
\end{equation}
Then, we formulate the following NLS:
\begin{equation}
\label{eq:icp_nls}
    \min \limits_{\hat{\bb{R}}_i,\hat{\bb{t}}_i} \sum^K_{j=1} ||\hat{\bb{s}}^1_j-\hat{\bb{R}}_i\hat{\bb{s}}^i_j-\hat{\bb{t}}_i||^2,
\end{equation}
where $\hat{\bb{R}}_i$ and $\hat{\bb{t}}_i$ are the $i$-th microphone array's rotation and translation to the first array, respectively. As it has been established in the literature \cite{ICP,icp_sovler}, there exists an analytical solution to Eq. (\ref{eq:icp_nls}):
\begin{align}
\label{eq:icp_solver}
    \bb{Q}_i&=\sum^K_{j=1}(\hat{\bb{s}}^1_j-\frac{1}{K}\sum^K_{k=1} \hat{\bb{s}}^1_k)(\hat{\bb{s}}^i_j-\frac{1}{K}\sum^K_{k=1} \hat{\bb{s}}^i_k)^T, \\
\label{eq:icp_solver2}
    \bb{Q}_i&=\bb{U}\bb{\Sigma}\bb{V}^T, \hat{\bb{R}}_i=\bb{U}\bb{V}^T,
\end{align}
where Eq. (\ref{eq:icp_solver2}) stands for the singular value decomposition of $\bb{Q}_i$. Note that in our previous work \cite{wang}, the position parameters and asynchronous parameters are estimated separately, and only DOA and odometer information were used to estimate position parameters. However, in this paper, the IVE here uses hybrid TDOA, DOA of the first array, and odometer measurement to jointly optimize the position and asynchronous parameters, thereby potentially improving the accuracy of orientation parameters. These points will be illustrated in the simulations and experiments later in the paper.

\subsubsection{Final Joint Optimization} 
Utilizing the estimated initial values obtained from the IVE, we construct the weighted NLS using hybrid TDOA, DOA, and odometer measurements, then solve it by the GN method to refine microphone array positions, orientations, time offsets, clock drift rates, and the sound event locations simultaneously. Denote the unknown parameters $\bb{x}=[\bb{x}_{mic},\bb{s}]^T$, measurements 
$\bb{z}=[\bb{t}^H, \bb{r}^D, \bb{m}]^T$ and measurement function $\bb{f}(\bb{x})=[\bb{T}^H,\bb{r},\Delta \bb{s}]^T$. The formulation of the final NLS is shown below: \\
\begin{equation}
    \label{NLS}
    \min_{\bb{x}} \ (\bb{f}(\bb{x})-\bb{z})^T\bb{W}(\bb{f}(\bb{x})-\bb{z}),
\end{equation}
where 
$\bb{W}=diag(w_{tdoa}\bb{I}_{2KN-K-N},w_{doa}\bb{I}_{3NK}, w_{odo}\bb{I}_{3K-3})$ 
and $w_{tdoa},w_{doa},w_{odo}$ are weighted parameters. Note that the derivative of the rotation matrix in DOA via Lie algebra is based on the left perturbation assumption \cite[Chp. 7]{stateEstRobotics}, and the details are not elaborated here.
 
\section{Simulations}
\subsection{Simulation Setup}
We design three trajectories for the motion of the sound source. The first trajectory has a space of $3m\times3m\times3m$ with eight sound events (Trajectory 1); the second trajectory has a space of $2m \times 6m\times 2m$ with ten sound events (Trajectory 2); and the third trajectory has a space of $4m \times 4m \times 2m$ with fourteen sound sources (Trajectory 3). The number of microphone arrays is five. The true and initial values of microphone array locations and sound source locations are randomly generated in the corresponding trajectory space. Similarly, the true and initial values of microphone array orientations are set to be arbitrary rotation vectors whose angles range from $-\pi$ to $\pi$. True values of $\lvert\tau_{i,1} \rvert \leq 0.1s $, $\lvert\delta{i} \rvert \leq 10^{-4}s $ and their initial values are set to zero. 

There are three levels of TDOA noises ($\sigma_{tdoa}= 0.05,0.1,0.5 ms$), and three levels of DOA noises ($\sigma_{doa}= 5,10,15^{\circ}$). In each level of TDOA noise and that of DOA noise, three trajectories are used, and in each trajectory, 200 Monto Carlo simulations are performed. Level of odometer measurement noises is set to be $\sigma_{odo}= 3\times10^{-2} m$. The results of three trajectories in each setting of measurement noises are combined for quantitative analysis. 

\begin{figure*}[htbp]
\centering

\includegraphics[width=0.9\linewidth]{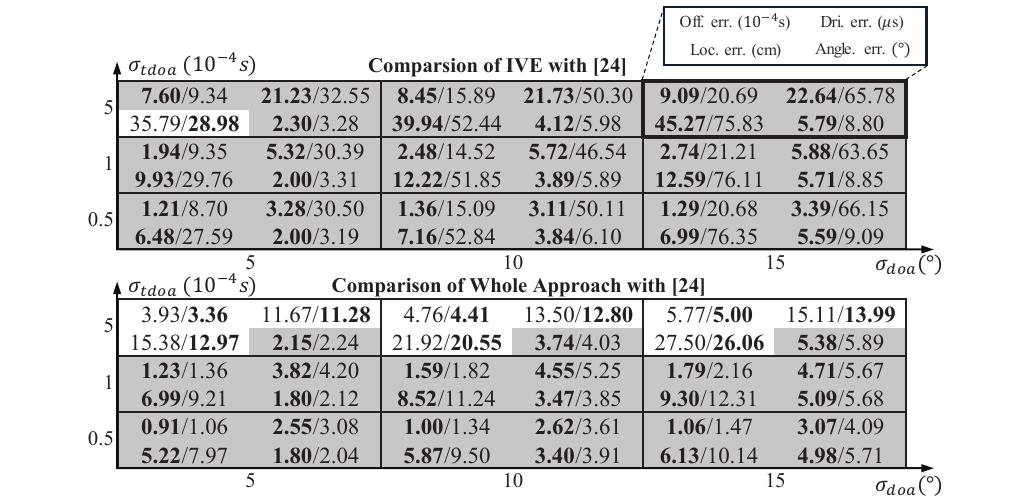} 
\caption{Comparison of simulation results between our IVE method/whole approach and \cite{wang}'s IVE method/whole approach under three levels TDOA noises ($\sigma_{tdoa}=0.05ms,0.1ms,0.5ms$) and three levels DOA noises ($\sigma_{doa}=0.05ms,0.1ms,0.5ms$). Four metrics are used: average RMSE of microphone array locations (Loc. err. (cm)), orientations (Angle err. (°)), time offsets (Off. err. ($10^{-4}s$)) and clock drift rates (Dri. err. ($\mu s$)). In this figure, ``A/B'' means that A is our IVE method or our whole approach, and B is that of \cite{wang}. Bold or gray-filled area means better.}
\label{simulationResult}
\end{figure*}

\subsection{Results and Discussions}
The estimation errors of the IVE method and final estimation errors for microphone array locations (Loc.err.), orientations (Angle.err.), time offsets (Off.err.), and clock drift rates (Dri.err.) are evaluated as the average root-mean-square errors (RMSE) within the interquartile range (IQR). These simulation results are shown in Fig. \ref{simulationResult}. The upper part of Fig. \ref{simulationResult}, titled `comparison of IVE Method with \cite{wang}', shows that our IVE method performs better than that of \cite{wang} in most cases, especially under relatively low TDOA noises. 

Also, our IVE method is less sensitive to DOA noises when estimating microphone array locations and two asynchronous timing parameters, while these parameters estimated by IVE method of \cite{wang} are influenced by DOA noises obviously because most steps in their IVE method using DOA only. These results indicate that our IVE method is more advantageous in cases of low or fair level of TDOA noises.

The lower part of Fig. \ref{simulationResult}, titled `comparison of whole approach with \cite{wang}', shows that our whole approach has better calibration accuracy under the moderate or low levels of TDOA noises. However, the performance of our method is worse than that of \cite{wang} when TDOA measurements contain large noises. The simulation results indicate that our method is more dependent on the noise levels of TDOA measurements.


\begin{figure}[htbp]
\centering
\includegraphics[scale=0.18]{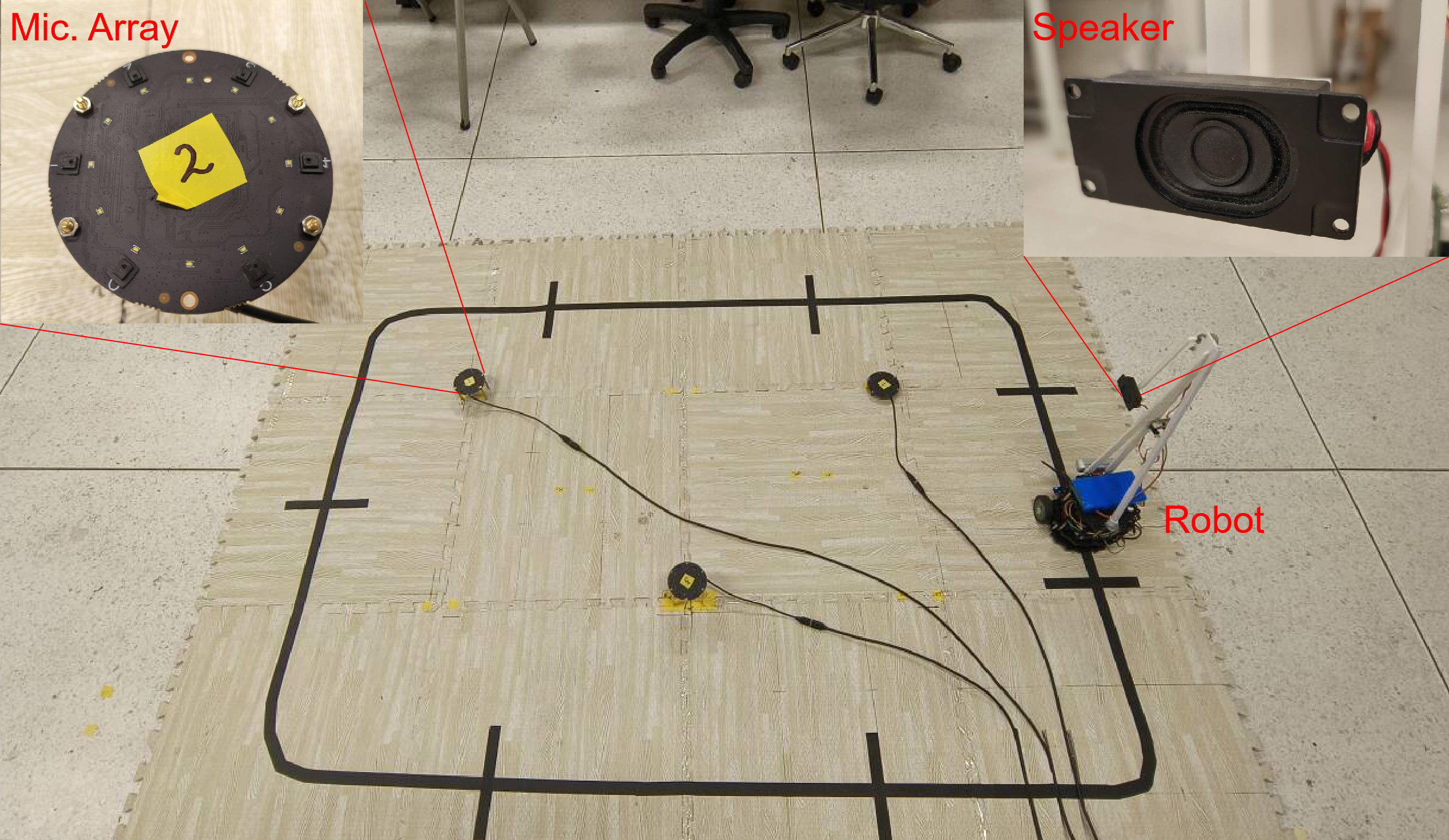}
\caption{The multiple microphone arrays calibration scenario for real-world experiments.}
\label{fig:realScene}
\end{figure}

\section{Real World Experiments}
For experiments, we use the real audio data in our previous work \cite{Hybrid-TDOA}. Note that the data is collected from three arrays distributed in an indoor environment as shown in Fig. \ref{fig:realScene}, which can be used to verify calibration methods for both the single array scenario and the multiple array scenario. From the dataset obtained in \cite{Hybrid-TDOA}, we calculate the inter-array TDOA-S, TDOA-M, and DOA measurements from 15 sets of calibration data with different true values of microphone array locations. From the ground truth values, we can estimate the noise levels of TDOA and DOA. Based on the available 15 sets of data, we found that the $\hat{\sigma}_{tdoa}$ is about $0.035$ms, and the average angle error of DOA is around $7.9$°. 

We compare the proposed approach with the method in \cite{wang}.
Because the true values of time offset and clock drift rates are unknown, we only evaluate the average RMSE of the microphone array position and rotation. The results of array location errors using the IVE method (Init. Loc. err. (cm)), array orientation errors using the IVE method (Init. Ori. err. (°)), array location errors using our approach (Loc. err. (cm)), and array orientation errors using our approach (Ori. err. (°)) are shown below, 
\begin{table}[htbp]
    \caption{Real-World Experiment Results (Bold means better)}
    \centering
    \resizebox{0.48\textwidth}{!}{
    \begin{tabular}{ccccc}
      \toprule
       & Init. Loc. err. & Init. Ori. err.  & Final  Loc. err. & Final Ori. err.\\
      \midrule
      Ours & 36.5 &  \textbf{5.35} &  \textbf{7.9} & \textbf{5.81}\\
      \cite{wang} & \textbf{15.2} & 7.95 & 9.3 & 6.28\\
      \bottomrule
    \end{tabular}
    }
\end{table}

It can be seen from Table I that the accuracy of our IVE is worse than the initialization method of \cite{wang}. This is due to the effects of relatively large TDOA noises ($\hat{\sigma}_{tdoa}$ is about $3\times10^{-4}s$). Despite the above, the final calibration performance of our approach is slightly better than that of \cite{wang}. If the noise level of TDOA gets larger, the performance of our method might become worse than that of \cite{wang}. These observations are consistent with the simulation results under large TDOA noises shown in Section IV.

\section{Conclusions}
By combining hybrid TDOA (TDOA-M and TDOA-S), DOA, and odometer measurements of a moving robot carrying a speaker, we have proposed a method for joint calibration of multiple asynchronous microphone arrays and sound source locations. Our proposed approach contains an IVE stage and the final optimization step. Our IVE method first constructs NLS using hybrid TDOA, DOA of the reference array, and odometer measurement to estimate microphone array positions, asynchronous parameters (time offsets, clock drift rates), and sound source locations. Then, given these estimation results and DOA measurements, we use the ICP method to estimate the orientations of microphone arrays. In the final optimization step, the initial values from the IVE are refined via NLS to achieve more accurate results. Extensive simulations and experiments show that for scenarios with low or moderate TDOA noise levels, our approach results in improved accuracy in comparison to SOTA methods. Future work will focus on enhancing the calibration performance of our proposed approach under large TDOA noises.
\bibliographystyle{IEEEtran}
\bibliography{ref}

\begin{thebibliography}{10}
\providecommand{\url}[1]{#1}
\csname url@samestyle\endcsname
\providecommand{\newblock}{\relax}
\providecommand{\bibinfo}[2]{#2}
\providecommand{\BIBentrySTDinterwordspacing}{\spaceskip=0pt\relax}
\providecommand{\BIBentryALTinterwordstretchfactor}{4}
\providecommand{\BIBentryALTinterwordspacing}{\spaceskip=\fontdimen2\font plus
\BIBentryALTinterwordstretchfactor\fontdimen3\font minus \fontdimen4\font\relax}
\providecommand{\BIBforeignlanguage}[2]{{%
\expandafter\ifx\csname l@#1\endcsname\relax
\typeout{** WARNING: IEEEtran.bst: No hyphenation pattern has been}%
\typeout{** loaded for the language `#1'. Using the pattern for}%
\typeout{** the default language instead.}%
\else
\language=\csname l@#1\endcsname
\fi
#2}}
\providecommand{\BIBdecl}{\relax}
\BIBdecl

\bibitem{app-track}
A.~Plinge and G.~A. Fink, ``Multi-speaker tracking using multiple distributed microphone arrays,'' in \emph{2014 IEEE International Conference on Acoustics, Speech and Signal Processing (ICASSP)}, 2014, pp. 614--618.

\bibitem{app-uav}
K.~Nakadai, M.~Kumon, H.~G. Okuno, K.~Hoshiba, M.~Wakabayashi, K.~Washizaki, T.~Ishiki, D.~Gabriel, Y.~Bando, T.~Morito, R.~Kojima, and O.~Sugiyama, ``Development of microphone-array-embedded uav for search and rescue task,'' in \emph{2017 IEEE/RSJ International Conference on Intelligent Robots and Systems (IROS)}, 2017, pp. 5985--5990.

\bibitem{evers2020locata}
C.~Evers, H.~W. L{\"o}llmann, H.~Mellmann, A.~Schmidt, H.~Barfuss, P.~A. Naylor, and W.~Kellermann, ``{The LOCATA challenge: Acoustic source localization and tracking},'' \emph{IEEE/ACM Transactions on Audio, Speech, and Language Processing}, vol.~28, pp. 1620--1643, 2020.

\bibitem{lagace2023ego}
P.-O. Lagac{\'e}, F.~Ferland, and F.~Grondin, ``Ego-noise reduction of a mobile robot using noise spatial covariance matrix learning and minimum variance distortionless response,'' in \emph{2023 IEEE/RSJ International Conference on Intelligent Robots and Systems (IROS)}, 2023, pp. 3533--3538.

\bibitem{Evers2018}
C.~Evers and P.~A. Naylor, ``Acoustic slam,'' \emph{IEEE/ACM Transactions on Audio, Speech, and Language Processing}, vol.~26, no.~9, p. 1484–1498, 2018.

\bibitem{hu2023}
Q.~Hu, N.~Ma, and G.~J. Brown, ``{Robust binaural sound localisation with temporal attention},'' in \emph{IEEE International Conference on Acoustics, Speech and Signal Processing (ICASSP)}, 2023, pp. 1--5.

\bibitem{Fu2024}
L.~Fu, Y.~He, J.~Wang, X.~Qiao, and H.~Kong, ``I-asm: Iterative acoustic scene mapping for enhanced robot auditory perception in complex indoor environments,'' in \emph{2024 IEEE/RSJ International Conference on Intelligent Robots and Systems (IROS)}, 2024, pp. 12\,318--12\,323.

\bibitem{Grondin}
P.~Gerstoft, Y.~Hu, M.~J. Bianco, C.~Patil, A.~Alegre, Y.~Freund, and F.~Grondin, ``Audio scene monitoring using redundant ad-hoc microphone array networks,'' \emph{IEEE Internet of Things Journal}, vol.~9, no.~6, p. 4259–4268, 2022.

\bibitem{soundLocReview}
C.~Rascon and I.~Meza, ``Localization of sound sources in robotics: A review,'' \emph{Robotics and Autonomous Systems}, vol.~96, pp. 184--210, 2017.

\bibitem{app-nakadai}
H.~G. Okuno and K.~Nakadai, ``Robot audition: Its rise and perspectives,'' in \emph{2015 IEEE International Conference on Acoustics, Speech and Signal Processing (ICASSP)}, 2015, pp. 5610--5614.

\bibitem{Molina}
A.~K. Katsaggelos, S.~Bahaadini, and R.~Molina, ``Audiovisual fusion: Challenges and new approaches,'' \emph{Proceedings of the IEEE}, vol. 103, no.~9, pp. 1635--1653, 2015.

\bibitem{Manocha2017}
M.~Ye, Y.~Zhang, D.~Manocha, and R.~Yang, ``3d reconstruction in the presence of glasses by acoustic and stereo fusion,'' \emph{IEEE Transactions on Pattern Analysis and Machine Intelligence}, vol.~40, no.~8, pp. 1785--1798, 2017.

\bibitem{qian2021multi}
X.~Qian, M.~Madhavi, Z.~Pan, J.~Wang, and H.~Li, ``{Multi-target DoA estimation with an audio-visual fusion mechanism},'' in \emph{IEEE International Conference on Acoustics, Speech and Signal Processing (ICASSP)}, 2021, pp. 4280--4284.

\bibitem{Verellen2020}
T.~Verellen, R.~Kerstens, D.~Laurijssen, and J.~Steckel, ``Urtis: A small 3d imaging sonar sensor for robotic applications,'' in \emph{ICASSP 2020 - 2020 IEEE International Conference on Acoustics, Speech and Signal Processing (ICASSP)}, 2020, pp. 4801--4805.

\bibitem{Ferreira}
Y.~R. Petillot, G.~Antonelli, G.~Casalino, and F.~Ferreira, ``Underwater robots: From remotely operated vehicles to intervention-autonomous underwater vehicles,'' \emph{IEEE Robotics \& Automation Magazine}, vol.~26, no.~2, pp. 94--101, 2019.

\bibitem{WangLin}
L.~Wang, T.-K. Hon, J.~D. Reiss, and A.~Cavallaro, ``Self-localization of ad-hoc arrays using time difference of arrivals,'' \emph{IEEE Transactions on Signal Processing}, vol.~64, no.~4, pp. 1018--1033, 2016.

\bibitem{SuKong}
D.~Su, H.~Kong, S.~Sukkarieh, and S.~Huang, ``Necessary and sufficient conditions for observability of slam-based tdoa sensor array calibration and source localization,'' \emph{IEEE Transactions on Robotics}, vol.~37, no.~5, pp. 1451--1468, 2021.

\bibitem{su2015simultaneous}
D.~Su, T.~Vidal-Calleja, and J.~V. Miro, ``Simultaneous asynchronous microphone array calibration and sound source localisation,'' in \emph{2015 IEEE/RSJ International Conference on Intelligent Robots and Systems (IROS)}, 2015, pp. 5561--5567.

\bibitem{su2020asynchronous}
------, ``Asynchronous microphone arrays calibration and sound source tracking,'' \emph{Autonomous Robots}, vol.~44, no.~2, pp. 183--204, 2020.

\bibitem{Li2024}
X.~Li, H.~Deng, J.~Wang, L.~Fu, and H.~Kong, ``Information-aware joint calibration of microphone array and sound source localization,'' in \emph{The 14th International Conference on Indoor Positioning \& Indoor Navigation (IPIN)}, 2024, pp. 1--6.

\bibitem{Ono}
N.~Ono, H.~Kohno, N.~Ito, and S.~Sagayama, ``Blind alignment of asynchronously recorded signals for distributed microphone array,'' in \emph{2009 IEEE Workshop on Applications of Signal Processing to Audio and Acoustics}, 2009, pp. 161--164.

\bibitem{Plinge2016}
A.~Plinge, F.~Jacob, R.~Haeb-Umbach, and G.~A. Fink, ``Acoustic microphone geometry calibration: An overview and experimental evaluation of state-of-the-art algorithms,'' \emph{IEEE Signal Processing Magazine}, vol.~33, no.~4, pp. 14--29, 2016.

\bibitem{2D-calib1}
A.~Plinge and G.~A. Fink, ``Geometry calibration of multiple microphone arrays in highly reverberant environments,'' in \emph{2014 14th International Workshop on Acoustic Signal Enhancement (IWAENC)}, 2014, pp. 243--247.

\bibitem{2D-calib2}
A.~Plinge, G.~A. Fink, and S.~Gannot, ``Passive online geometry calibration of acoustic sensor networks,'' \emph{IEEE Signal Processing Letters}, vol.~24, no.~3, pp. 324--328, 2017.

\bibitem{2D-Calib3}
D.~Hu, Z.~Chen, and F.~Yin, ``Passive geometry calibration for microphone arrays based on distributed damped newton optimization,'' \emph{IEEE/ACM Transactions on Audio, Speech, and Language Processing}, vol.~29, pp. 118--131, 2021.

\bibitem{3D-Calib1}
R.~Wang, Z.~Chen, and F.~Yin, ``Doa-based three-dimensional node geometry calibration in acoustic sensor networks and its cramér–rao bound and sensitivity analysis,'' \emph{IEEE/ACM Transactions on Audio, Speech, and Language Processing}, vol.~27, no.~9, pp. 1455--1468, 2019.

\bibitem{3D-Calib2}
S.~Woźniak and K.~Kowalczyk, ``Passive joint localization and synchronization of distributed microphone arrays,'' \emph{IEEE Signal Processing Letters}, vol.~26, no.~2, pp. 292--296, 2019.

\bibitem{3D-Calib3}
C.~Sugiyama, K.~Itoyama, K.~Nishida, and K.~Nakadai, ``Assessment of simultaneous calibration for positions, orientations, and time offsets in multiple microphone arrays systems,'' in \emph{2023 IEEE/SICE International Symposium on System Integration (SII)}, 2023, pp. 1--6.

\bibitem{wang}
J.~Wang, Y.~He, D.~Su, K.~Itoyama, K.~Nakadai, J.~Wu, S.~Huang, Y.~Li, and H.~Kong, ``{SLAM-based joint calibration of multiple asynchronous microphone arrays and sound source localization},'' \emph{IEEE Transactions on Robotics}, 2024, {DOI: 10.1109/TRO.2024.3410456}.

\bibitem{Hybrid-TDOA}
C.~Zhang, J.~Wang, and H.~Kong, ``Asynchronous microphone array calibration using hybrid tdoa information,'' in \emph{2024 IEEE/RSJ International Conference on Intelligent Robots and Systems (IROS)}, 2024, pp. 913--918.

\bibitem{stateEstRobotics}
T.~D. Barfoot, \emph{State Estimation for Robotics}, 2nd~ed.\hskip 1em plus 0.5em minus 0.4em\relax Cambridge University Press, 2024.

\bibitem{lieGroup}
A.~Kirillov~Jr, \emph{An Introduction to Lie Groups and Lie Algebras}, ser. Cambridge Studies in Advanced Mathematics.\hskip 1em plus 0.5em minus 0.4em\relax Cambridge University Press, 2008.

\bibitem{ICP}
F.~Pomerleau, F.~Colas, and R.~Siegwart, ``A review of point cloud registration algorithms for mobile robotics,'' \emph{Foundations and Trends in Robotics}, vol.~4, no.~1, pp. 1--104, 2015.

\bibitem{icp_sovler}
K.~S. Arun, T.~S. Huang, and S.~D. Blostein, ``Least-squares fitting of two 3-d point sets,'' \emph{IEEE Transactions on Pattern Analysis and Machine Intelligence}, vol. PAMI-9, no.~5, pp. 698--700, 1987.

\end{thebibliography}

\end{document}